\documentclass[aps,prl,twocolumn,showpacs,floatfix,superscriptaddress]{revtex4}
\usepackage{amsmath,amssymb,bm,
graphicx,times}

\newcommand{\be}{\begin{equation}}
\newcommand{\ee}{\end{equation}}
\newcommand{\bea}{\begin{eqnarray}}
\newcommand{\eea}{\end{eqnarray}}
\newcommand{\bw}{\begin{widetext}}
\newcommand{\ew}{\end{widetext}}
\newcommand{\ba}{\beta}

\newcommand{\vvr}{\bm r} 
\newcommand{\vvv}{\bm v}

\newcommand{\kommentar}[1]{}

\begin{document}
 
\title{Thermodynamic formalism for the Lorentz gas with open boundaries in
$d$ dimensions}
\author{Henk van Beijeren}
\email{H.vanBeijeren@phys.uu.nl}
\affiliation{Institute for Theoretical Physics, Utrecht University,
Leuvenlaan 4, 3584 CE Utrecht, The Netherlands} 
\author{Oliver M\"ulken}
\email{oliver.muelken@physik.uni-freiburg.de}
\affiliation{Theoretical polymer physics, University of Freiburg,
Hermann-Herder-Stra{\ss}e 3a, 79104 Freiburg i.\ Br.\, Germany}
\affiliation{Institute for Theoretical Physics, Utrecht University,
Leuvenlaan 4, 3584 CE Utrecht, The Netherlands}
\date{\today} 
\begin{abstract}
A Lorentz gas may be defined as a system of fixed dispersing scatterers,
with a single light particle moving among these and making specular
collisions on encounters with the scatterers. For a dilute Lorentz gas
with open boundaries in $d$ dimensions we relate the thermodynamic
formalism to a random flight problem. Using this representation we
analytically calculate the central quantity within this formalism, the
topological pressure, as a function of system size and a temperature-like
parameter $\ba$. The topological pressure is given as the sum of the
topological pressure for the closed system and a diffusion term with a
$\ba$-dependent diffusion coefficient.  From the topological pressure we
obtain the Kolmogorov-Sinai entropy on the repeller, the topological
entropy, and the partial information dimension.  
\end{abstract}
\maketitle

\section{Introduction}

In the seventies of the last century Sinai, Ruelle, and Bowen developed a
formalism for calculating dynamical properties of chaotic dynamical
systems \cite{sinai,ruelle,bowen}.  This formalism resembles very much
Gibbs ensemble theory and therefore was given the name {\sl thermodynamic
formalism}. From a partition function defined as an integral over phase
space of a particular weight function, one may derive several dynamical
quantities, such as the Kolmogorov-Sinai (KS) entropy, the topological
entropy, the escape rate, or the partial information dimension
\cite{er1985,beck}.

Although the power of this formalism has been widely recognized, there are
only few examples so far of physically interesting systems for which the
topological pressure and related functions could be evaluated explicitly.
One such system is the dilute random Lorentz gas. Generally spoken a
Lorentz gas is a system consisting of fixed dispersing scatterers, among
which a single light particle (or alternatively a cloud of mutually
non-interacting light particles) moves about, making specular collisions
on each encounter with a scatterer. The scatterers may either be placed on
a periodic lattice or at random positions in space. Usually the scatterers
are taken to be disks, in $d=2$, spheres, in $d=3$, or hyperspheres, for
$d>3$. If these are placed on a periodic lattice the resulting system is
also known as the Sinai billiard~\cite{SB}. In fact this is also the model
Lorentz~\cite{lorentz} had in mind originally for describing the transport
of conduction electrons in metals (still in the pre-quantum era). However
the kinetic equation he proposed to describe this system, presently known
as the Lorentz-Boltzmann equation (see for instance
\cite{Hauge,dorfman,vb1982}), in fact is more appropriate for the model
with random scatterer locations, at low density of scatterers. The
Lorentz-Boltzmann equation and some of its generalizations to higher
densities~\cite{vLW,WvL,EW} allow for analytic calculations of transport
coefficients and other fundamental nonequilibrium properties.

Recently, several dynamical properties have been calculated analytically
for open and closed random Lorentz gases by using an extended
Lorentz-Boltzmann equation approach \cite{vbd1995,
vbld,vbld2000,dwvb2004}.  Van Beijeren and Dorfman~\cite{vbd}
alternatively used the thermodynamic formalism to calculate the
topological pressure and related quantities for a dilute d-dimensional
random Lorentz gas in equilibrium. For 2 dimensions this was extended by
the present authors~\cite{mvb2004} to the case of a dilute random Lorentz
gas under the combined actions of a driving field and an isokinetic
gaussian thermostat. 

\section{thermodynamic formalism}

In chaos theory a central role is played by the time evolution of
infinitesimal volumes in phase space. For chaotic systems such a volume
element will grow in some directions and shrink in others. The factor by
which the projection of the volume element onto its unstable (expanding)
manifold increases over a time $t$ is called the {\em stretching factor}.
For an infinitesimal volume centered originally around the phase point
$(\vvr,\vvv)$ we will denote its value as $\Lambda(\vvr,\vvv,t)$.
Similarly, the contraction factor is the factor by which the projection
onto the stable (contracting) manifold decreases over time $t$. In analogy
to the Gibbs ensemble formalism of statistical mechanics, we may define a
{\sl dynamical partition function} $Z(\ba,t)$ as a weighted integral over
phase space. For closed systems, indicated by a subscript $0$, points in
phase space are weighted by powers of their stretching factor, according
to
\be
Z_0(\ba,t) = \int d\mu(\vvr,\vvv) \left[ \Lambda(\vvr,\vvv,t)
\right]^{1-\ba},
\label{partfunct0}
\ee
where the integration is over an appropriate stationary measure. In
systems with escape, phase space trajectories are removed from the
ensemble if they hit an absorbing boundary. In this case the definition of
the dynamical partition function has to be generalized to
\be
Z(\ba,t) = \int d\mu(\vvr,\vvv) \left[ \Lambda(\vvr,\vvv,t)
\right]^{1-\ba}\chi_{tr}(\vvr,\vvv,t) ,
\label{partfunct}
\ee
with $\chi_{tr}(\vvr,\vvv,t)=0$ if the trajectory starting from
$(\vvr,\vvv)$ at time 0 hits the absorbing boundary before time $t$ and
$\chi_{tr}(\vvr,\vvv,t)=1$  otherwise. In our analogy to statistical
mechanics the parameter $\ba$ behaves like an inverse temperature and as
the analogon of the Helmholtz free energy we obtain the {\em topological
pressure} $P(\ba)$ as
\be
P(\ba) = \lim_{t\to\infty} \frac{1}{t} \ln Z(\ba,t).
\label{toppress}
\ee

The {\sl dynamical entropy function} $h(\ba)$ is given by the Legendre
transform of $P(\ba)$ as,
\be
h(\ba) = P(\ba) - \ba P'(\ba),
\label{dynentr}
\ee
where $P'(\ba) = d P(\ba)/ d \ba$.  For special values of $\ba$ the
dynamical entropy function can be identified with dynamical properties,
because for long times we have $\Lambda \simeq \exp(t\sum^+_i\lambda_i)$,
where $\sum^+_i\lambda_i$ is the sum of all positive Lyapunov exponents
$\lambda_i$.  Specifically, $h(\ba)$ equals the topological entropy
$h_{\rm top}$ for $\ba=0$ and the KS entropy $h_{KS}$ for $\ba=1$. If the
system is closed, the KS entropy equals the sum of positive Lyapunov
exponents because $P_0( 1)=0$, as follows directly from Eqs.\
(\ref{partfunct0}) and (\ref{toppress}).

However, for open systems $P( 1)=-\gamma$, where $\gamma$ is the
asymptotic escape rate. The relationship $P'(1) = \sum^+_i\lambda_i$ still
holds, but now the Lyapunov exponents are defined on the {\em repeller},
i.e.\ the subset of phase space from which no trajectories escape. The
point where $P(\ba)$ intersects the $\ba$-axis can be related to the
partial Hausdorff dimension, which is a fractal dimension of a line across
the stable manifold of the attractor.  Another fractal dimension
associated with the topological pressure is the partial information
dimension which is given by the intersection point with the $\ba$-axis of
the tangent to $P(\ba)$ at $P(1)$.  For the closed system both dimensions
coincide and are equal to $1$.

\section{Dynamical partition function for the open Lorentz gas in $d$
dimensions}

In this section we will calculate $Z(\ba,t)$ for a dilute random open
Lorentz gas in $d$ dimensions, with $N$ fixed (hyper)spherical scatterers
of radius $a$ distributed randomly inside a finite volume $V$.  In
addition there is one point particle moving with velocity $v$ along
straight lines between specular collisions with the scatterers. If the
particle leaves $V$ it escapes. "Dilute" implies the condition $na^d\ll
1$, with the density $n=N/V$.

\begin{figure}
\centerline{\includegraphics[width=0.75\columnwidth]{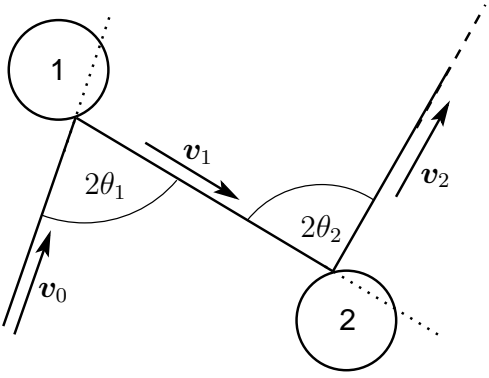}}
\caption{Sketch of a trajectory in two dimensions of the light particle
starting form the initial point and following the first two collisions.
}\label{colls}
\end{figure}

The condition of diluteness allows us to approximate the dynamics of the
light particle as a random flight, in which each trajectory between
subsequent collisions is sampled from an exponential distribution of free
path lengths and the collision parameters of each collision are sampled
from a distribution corresponding to a homogeneous bundle hitting the
scatterer. In other words: all effects resulting from recollisions are
completely ignored. Under these approximations the dynamical partition
function may be rewritten as
\begin{widetext}
\bea
Z(\ba,t) &=&e^{P_0(\ba) t}\int d\vvr\int d\vvv\delta(|\vvv|-v_0)\sum_{l=0}^\infty 
\int_0^\infty dt_1\cdots
dt_l \int d\mu(\hat{\sigma}_1)\cdots d\mu(\hat{\sigma}_l)\nonumber\\
&&\Theta\left(t-\sum_{i=1}^l
t_i\right)
\exp\left[-\left((\nu_d+P_0(\ba))(t-\sum_{i=1}^lt_i)\right)\right]
\chi\left(\vvr_l+\vvv_l'(t-\sum_{i=1}^lt_i)\right)
\prod_{j=1}^l W(\hat{\sigma}_j,t_j)\chi(\vvr_j).
\label{Zrandomflight}
\eea
\end{widetext}
Here $P_0$ is the topological pressure for the closed Lorentz gas at the
same density in equilibrium; $\hat{\sigma}_i$ denotes the collision normal
at the $i$th collision and $d\mu(\hat\sigma)$ denotes the probability
measure for scattering with collision normal $\hat\sigma$; $\Theta(x)$
denotes the unit step function, i.e.\ $\Theta(x)= 1$ for $x\ge 0$ and
$\Theta(x)= 0$ for $x<0$; $\nu_d \equiv1/\bar{\tau}_d$ is the average
collision rate, given for dimension $d$ as \cite{vbd} 
\be 
\nu_d = \frac{2 n v a^{d-1}
\pi^{\frac{d-1}{2}}}{(d-1) \Gamma\left(\frac{d-1}{2}\right)},
\label{nud}
\ee
$\vvr_j$ is the position of the light particle at the $j$th collision, and
$\chi(\vvr)$ is the characteristic function satisfying $\chi(\vvr)=1$ for
$\vvr$ inside $V$ and 0 otherwise. We implicitly assumed that the boundary
is not concave, so if both $\chi(\vvr_j)$ and $\chi(\vvr_{j+1})$ equal
unity the same is true for the characteristic function of all points in
between. In addition, $\vvv_l'$ is the velocity of the light particle just
after the $l$th collision. Finally, $W(\hat{\sigma},t)$ is an effective
free flight transition rate, defined as
\be
W(\hat{\sigma},t)=\nu_d 
e^{-( P_0(\ba)+\nu_d) t}
\left[\Lambda_d(\theta,t)\right]^{1-\ba}.
\label{flightdistr}
\ee
Here the stretching factor is given by \cite{vbld,dwvb2004}
\be
\Lambda_d(\theta,t) = \left(\frac{2vt}{a}\right)^{d-1}(\cos\theta)^{d-3},
\label{stretch}
\ee
where $\theta$ is the scattering angle, defined through
$\cos\theta=-\hat{v}\cdot\hat{\sigma}$, with $\hat{v}$ the unit vector
along the velocity of the light particle before the collision, see Fig.\
\ref{colls}.
Since in Eq.\ (\ref{flightdistr}) the only dependence of the integrand on
$\hat{\sigma_j}$ occurs through $\cos \theta_j$, the integrations over
$d\mu(\hat{\sigma_j})$ may be reduced as 
\[
\int d\mu(\hat{\sigma_j})\to
(d-1)\int_0^{\pi/2}d\theta_j
\cos\theta_j\sin^{d-2}\theta_j.
\] 
Note that in Eq.\ (\ref{flightdistr}) the actual free flight distribution
has been changed to an effective free flight distribution by
multiplication by the $(1-\ba)$th power of the stretching factor and by
the factor $\exp(-P_0(\ba)t)$. Similarly, the distribution of collision
normals has been changed to an effective distribution. Indeed, after this
rearrangement the integral of $W(\hat{\sigma},t)$ over
$d\mu(\hat{\sigma})$ and positive $t$ equals unity~\cite{vbd}.  The second
moments of both time and displacement for the effective distribution
$W(\hat{\sigma},t)$ are well-defined, and the resulting effective random
flight process for given initial direction gives rise to a convergent
average displacement after $n$ steps in the limit $n\to \infty$.
Therefore, on large time and length scales it leads to a normal diffusion
process, with a $\ba$-dependent diffusion coefficient $D(\beta)$.

On division of the logarithm of Eq.\ (\ref{Zrandomflight}) by $t$ the
first factor just simply reduces to $P_0(\ba)$. The contribution from the
remaining factor may be interpreted, up to a normalization factor, as the
average survival probability of an initially homogeneously distributed
cloud of light particles of equal energy, under the effective random
flight process with absorbing boundaries described above. Since on large
time and length scales this becomes an effective diffusion process, this
survival probability will behave as $\exp(-k_0^2 D(\ba)t)$ for long times,
with $k_0$ determined by the geometry of the system and to some extent by
the boundary conditions on the random flight process.  Therefore the
second contribution is of the form $-k_0^2 D(\ba)$.  Combination of these
two brings us to the main result of this paper: the topological pressure
for the dilute open random Lorentz gas may be expressed as
\be
P(\ba)=P_0(\ba)-k_0^2D(\ba).
\label{pbeta}
\ee

What remains to be done now is finding an explicit expression for $D(\ba)$
as a function of both $\ba$ and $d$. This task is not particularly hard.
The Green-Kubo expression for the diffusion coefficient as a time integral
of the velocity autocorrelation function gives rise to
\bea
D(\ba) &=& \frac{1}{d} \int_0^\infty dt \ \langle \vec v(0)\cdot \vec v(t)
\rangle \nonumber \\
&=& \frac {v^2} d \left[\langle\tau_0\rangle+ \sum_{l=1}^\infty
\langle\cos(\pi-2\theta)\rangle^l \langle\tau\rangle\right].
\label{greenkubo}
\eea
The first term is the contribution from free flights from the initial time
until the first collision, the next terms result from free flights between
the $l$th and $(l+1)$th collisions. Since the collision cross-sections are
isotropic and all collisions are uncorrelated, the average direction of
the velocity after the $l$th collision is proportional to
$\langle\cos(\pi-2\theta)\rangle^l$.  Then we obtain from Eqs.\
(\ref{Zrandomflight}) and (\ref{flightdistr}) the time averages
\bea
\langle \tau_0 \rangle &=& \frac{\langle \tau^2 \rangle}{2 \langle \tau
\rangle} = \frac{
1}{\nu_d + P_0(\ba)} \left[1 + \frac{(d-1)(1-\ba)}{2}
\right] \label{tau0},\\
\langle \tau \rangle &=&  \frac{
1}{\nu_d + P_0(\ba)} \left[1 +
(d-1)(1-\ba)\right]
\label{tau}, 
\eea
together with the angular average
\be
\langle \cos(\pi-2\theta) \rangle = \frac{
-\ba(3-d)}{2(d-1
)+\ba(3-d)
}.
\label{anglaverage}
\ee
To understand the first equality in Eq.~(\ref{tau0}) one should realize
that the probability of finding the initial time on a free-time stretch of
length $\tau$ is proportional to $\tau$ and that the average time until
the first collision under these conditions is just $\tau/2$.  Furthermore
Eqs.\ (\ref{tau0}) and (\ref{tau}) only make sense for $\ba<d/(d-1)$ since
for larger $\ba$-values the effective free-time distribution shows a
divergence, see Eqs.\ (\ref{flightdistr}) and (\ref{stretch}).  Similarly
Eq.\ (\ref{anglaverage}) has to be restricted to the range $\ba> -1$, for
$d=2$, respectively $\ba< (d-1)/(d-3)$ for $d\ge 4$ (for $d=3$ there are
no restrictions).

The topological pressure for the closed Lorentz gas in equilibrium is
obtained from the first pole of the Laplace transform of the dynamical
partition function \cite{vbd,mvb2004} and for the closed Lorentz gas in
equilibrium it reads
\bea
P_0(\ba) &=&\nu_d\left\{ \left[
\frac{d-1}{2}
\left(\frac{2v}{a
\nu_d}\right)^{(d-1)(1-\ba)}\Gamma\left(d+\ba-d\ba\right)
\right.\right.
\nonumber \\
&& \times
\left.\left.
\frac{\Gamma\left(\frac{d-1}{2}\right)\Gamma\left(\frac{d-1+\ba(3-d)}{2}\right)}{\Gamma\left(d-1+\frac{\ba(3-d)}{2}\right)}
\right]^{\frac{1}{d+\ba-d\ba}} - 
1\right\}.\
\eea 
Note that for the closed Lorentz gas in equilibrium the topological
pressure vanishes for $\ba=1$.  Now $D(\ba)$ is given by
\bea
D(\ba) &=& \frac{v^2}{d(\nu_d+P_0(\ba))}\left[1 + \frac{(d-1)(1-\ba)}{2}
\right. \nonumber \\ 
&& \left. 
- \frac{[1 + (d-1)(1-\ba)] \ba(3-d)}{2
[d-1+\ba(3-d)
]} \right].
\label{Dba}
\eea
From this we can easily get the {\sl normal} diffusion coefficient by
setting $\ba=1$, i.e.,
\be
D(1) = \frac{v^2(d+1)}{4d\nu_d}.
\ee

\section{Other dynamical properties}

As mentioned before, from the dynamical partition function we may obtain
several other dynamical characteristics of our system.  For $\ba=1$ the
topological pressure equals minus the escape rate $\gamma$, thus we have
$\gamma=k_0^2 D(1)$.  From Eqs.~(\ref{dynentr}) and (\ref{pbeta}) we
obtain the dynamical entropy function $h(\ba)$ as
\be
h(\ba) = h_0(\ba) - k_0^2\left[ D(\ba) - \ba D'(\ba)
\right],
\label{hbeta}
\ee
with $D'(\ba) = d D(\ba)/d \ba$ and $h_0(\ba)$ the entropy function for
the closed Lorentz gas in equilibrium.

\begin{figure}
\centerline{\includegraphics[width=0.75\columnwidth]{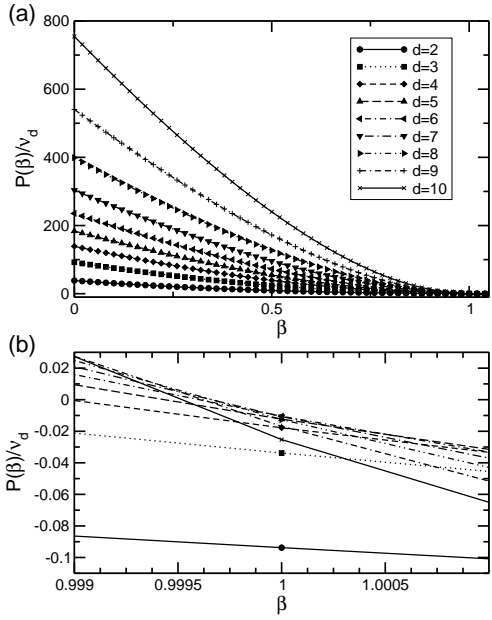}}
\caption{(a) Topological pressure divided by the collision frequency as a
function of $\ba$ for different dimensions $d$, with the parameter choice
$a=1$, $v=1$, $n=0.001$, and $k_0=0.001$. (b) A close-up of this for
$\ba$-values around 1.
}\label{press_beta}
\end{figure}

\begin{figure}
\centerline{\includegraphics[width=0.75\columnwidth]{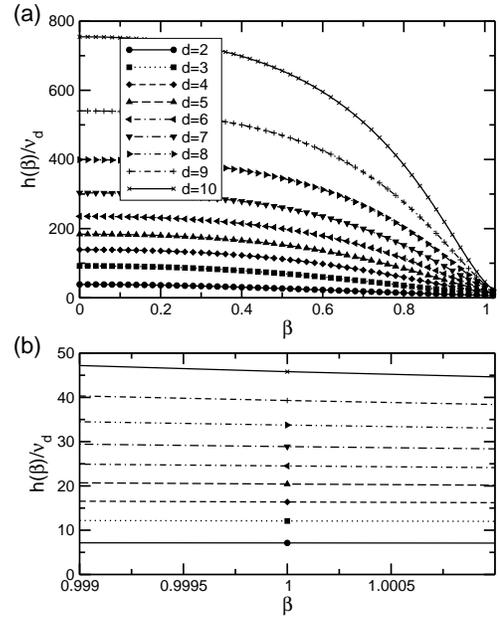}}
\caption{(a) Dynamical entropy divided by the collision frequency, as
function of $\ba$ for different dimensions $d$ and with the parameter
choice $a=1$, $v=1$, $n=0.001$, and $k_0=0.001$. (b) A close-up of this
for $\ba$-values around 1.
}\label{entr_beta}
\end{figure}

Figures \ref{press_beta} and \ref{entr_beta} show the topological pressure
and the dynamical entropy, respectively, divided by the collision
frequency, as functions of $\ba$ for different dimensions $d$. As
expected, $P(\ba)$ is negative for $\ba=1$ because $P(1) = -\gamma$, see
the inset of Fig.\ \ref{press_beta}. Furthermore, $P(\ba)$ is a convex
function for all dimensions considered. Since $k_0 \ll 1$ for large
systems, the deviations of $P(\ba)$ and $h(\ba)$ from their equilibrium
values are small.

\begin{figure}
\centerline{\includegraphics[width=0.85\columnwidth]{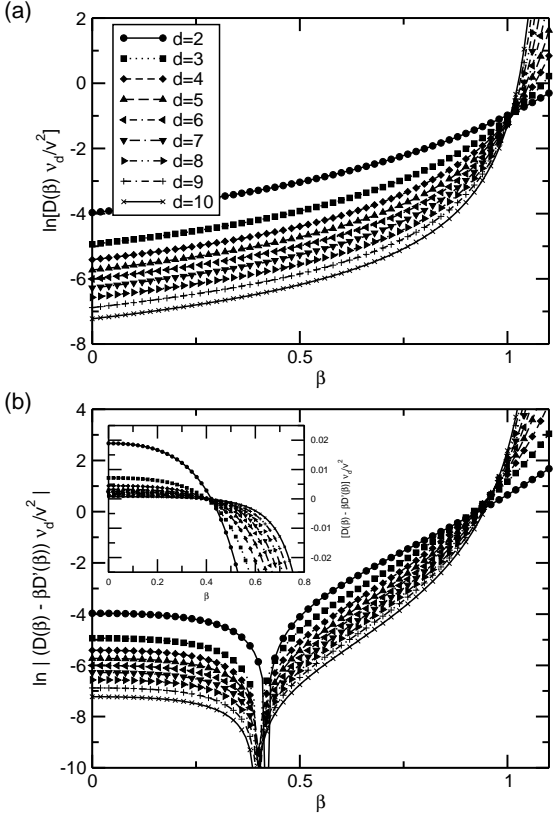}}
\caption{
Natural logarithms of the correction terms proportional to $k_0^2$ of (a)
the topological pressure, see Eq.\ (\ref{pbeta}), and (b) the absolute
value of the dynamical entropy, see Eq.\ (\ref{hbeta}), as a function of
$\ba$ for different dimensions $d$.  The inset in (b) shows a close-up of
$\left[ D(\ba) - \ba D'(\ba)\right]$ around $ \ba=0.4$, where this
function crosses zero.  All results are for $a=1$, $v=1$, and $n=0.001$
}\label{press_entr}
\end{figure}

The logarithms of the prefactors of the correction terms in Eqs.\
(\ref{pbeta}) and (\ref{hbeta}) proportional to $k_0^2$ are plotted in
Fig.\ \ref{press_entr}. For $P(\ba)$ this prefactor is $D(\ba)$, which is
always positive within the allowed ranges of $\ba$ and $d$. As can be seen
from Eq.\ (\ref{Dba}) $D(\ba)$ diverges at a pole located at
$\ba=(d-1)/(d-3)$.  In Fig.\ \ref{press_entr}(b) the logarithm of the
absolute value of the correction term for the dynamical entropy is plotted
because the prefactor changes sign at $\ba\approx0.4$, see inset Fig.\
\ref{press_entr}(b). Thus the correction to the topological entropy at
$\ba=0$ will be negative while the one for the KS-entropy at $\ba=1$ will
be positive (see also the discussion).

The KS entropy for general $d$ for the open Lorentz gas is given by
\be
h(1) = h_{KS} = h^0_{KS} - k_0^2 D(1) \left[ \frac{d+1}{2} -
\frac{h^0_{KS}}{\nu_d} \right], 
\label{hks}
\ee
where $h^0_{KS}$ is the equilibrium value of the KS entropy for the closed
Lorentz gas in $d$ dimensions.  The specific form of this as a function of
$d$ reads
\bea
h^0_{KS} &=& \nu_d(1-d) \left[{\cal C} -
\ln\left(\frac{2v}{a\nu_d}\right)\right] \nonumber \\
&& + \frac{\nu_d}{2}(3-d) \left[{\cal C}
+ \Psi\left(\frac{d+1}{2}\right)\right],
\label{hks0}
\eea
with ${\cal C}$ Euler's constant and $\Psi(x)$ the digamma function
\cite{gradshteyn}.

For $d=2$ and $d=3$ we can compare our results to previous calculations
based on an extended Lorentz-Boltzmann equation approach \cite{vbld2000}.
From Eq.\ (\ref{hks0}) we get the KS entropy for $d=2$ as
\be
h_{KS} = h^0_{KS} - k_0^2 \frac{3v^2}{8
\nu_2} \left[ \frac{3}{2} -
\frac{h^0_{KS}}{
\nu_2} \right] .
\label{hks2}
\ee
$h^0_{KS}$ is given by the one positive Lyapunov exponent in equilibrium
\be
h^0_{KS} = \lambda_0^+ = 
\nu_2 \left[ 1 -{\cal C} -
\ln\left(\frac{a
\nu_2}{v}\right)\right].
\label{hks20}
\ee
The KS entropy for $d=3$ follows as
\be
h_{KS} = h^0_{KS} - k_0^2\frac{v^2}{3
\nu_3} \left[ 2 -
\frac{h^0_{KS}}{
\nu_3} \right].
\label{hks3}
\ee
Here, $h^0_{KS}$ is given by the sum of the positive Lyapunov exponents in
equilibrium
\be
h^0_{KS} = \sum_i\lambda^+_{0,i} = 2
\nu_3 \left[\ln\left(\frac{2v}{a
\nu_3}\right)
- {\cal C}\right].
\label{hks30}
\ee 
As one should expect, Eqs.\ (\ref{hks2}-\ref{hks30}) are in agreement with
the previous calculation. 

From the escape rate formalism \cite{gn1990} we know that the KS entropy
equals the sum of positive Lyapunov exponents minus the escape rate. Note
that here the Lyapunov exponents are defined on the repeller. Furthermore,
for $\ba=1$ the topological pressure equals the escape rate $\gamma$ and
this can be expressed for the open Lorentz gas as $\gamma = k_0^2 D(1)$.
Therefore, we have an equation for the sum of positive Lyapunov exponents
of the open Lorentz gas,
\bea
\sum_i \lambda_i^+ &=&  \sum_i \lambda_{0,i}^+ +  k_0^2 D'(1)
\\
&=& h^0_{KS} 
+ k_0^2 D(1) \left[\frac
{1-d}{2} 
+ \frac{h^0_{KS}}{\nu_d} \right].
\label{lyapopen}
\eea
The correction term proportional to the diffusion coefficient is always
positive, because for low densities the term $\ln(2v/a\nu_d) \gg1$ in Eq.\
(\ref{hks30}) dominates. Hence, the sum of positive Lyapunov exponents is
always larger for the open system than for the corresponding closed
system. This again is in quantitative agreement with previous results
\cite{vbld2000}.

Since we have an expression for general $\ba$ we can also calculate the
topological entropy $h_{\rm top}$ which is given by $h(\ba)$ for $\ba=0$.
For general $d$ this is given by
\be
h(0) = h_{\rm top} = h^0_{\rm top} - k_0^2 \frac{2 D(1)
\nu_d}{h^0_{\rm top} +
\nu_d},  
\ee
with the equilibrium value of the topological entropy
\be
h_{\rm top}^0
= \nu_d \left\{ \left[ \frac{d-1}{\sqrt{2}} \left( \frac{2v}{a\nu_d} 
\right)^{\frac{d-1}{2}} \Gamma\left(\frac{d-1}{2}\right) 
\right]^{\frac{2}{d}} - 1 \right\}.
\ee
So we may conclude that the topological entropy is always smaller than in
equilibrium.

\begin{figure}
\centerline{\includegraphics[width=0.75\columnwidth]{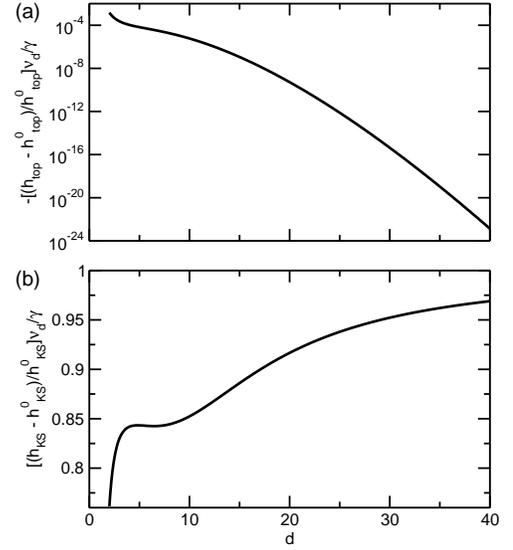}}
\caption{
Relative corrections to the topological and to the KS entropy. (a) shows
the natural logarithm of $(h_{\rm top}-h^0_{\rm top})\nu_d/(\gamma
h^0_{\rm top})$ and
(b) shows  $(h_{KS}-h^0_{KS})\nu_d/(\gamma h^0_{KS})$, both as functions
of $d$, for parameter values $a=1$, $v=1$ and $n=0.001$.
}\label{htop_hks}
\end{figure}

In Fig.\ \ref{htop_hks} the relative corrections to the KS entropy and the
topological entropy are plotted as functions of $d$.  For $h_{\rm top}$
the ratio to the equilibrium value is very small. It decreases
exponentially with $d$, because asymptotically for large $d$ $h^0_{top}$
is independent of the radius $a$, whereas the collision frequency $\nu_d$
is proportional to $a^(d-1)$.  For $h_{KS}$ the correction, scaled as in
the figure, appoaches unity in the limit of large $d$, as follows readily
from eqs.\ (\ref{hks}) and (\ref{nud}) plus Stirling's approximation
$\ln\Gamma(x)\approx x\ln x -x$.

More dynamical properties can be obtained from $P(\ba)$. The partial
Hausdorff dimension is given by the value of $\ba$ where $P(\ba)$
intersects the $\ba$-axis, whereas the partial information dimension is
given by the value of $\ba$ where the tangent at $P(1)$ intersects the
$\ba$-axis \cite{er1985,br1987}. For the latter we easily find that the
tangent is given by
\bea
&&P(1) - (1-\ba) P'(1) \nonumber\\
&& = (1-\ba) h^0_{KS} - k_0^2 \left[ D(1) - (1-\ba)
D'(1) \right].
\label{ptan}
\eea
Thus, the partial information dimension follows as
\bea
d_I &=& 1 - \frac{k_0^2 D(1)}{h^0_{KS} + k_0^2 D'(1)
} \nonumber \\
&=& 1- \frac{1}{\frac{1-d}{2}+h^0_{KS}(1/\gamma+1/\nu_d)}
\label{infdim}
\eea
and is clearly smaller than unity. For a large system with fixed density
an expansion for small escape rates $\gamma$ gives $d_I \approx 1 -
\gamma/h^0_{KS}$. The complexity of $P(\ba)$ and the $\ba$-dependence of
the diffusion coefficient prevent a calculation of the partial Hausdorff
dimension. However, for large systems, where $k_0$ is very small, the
partial Hausdorff dimension will be very close to $d_I$ because the point
of intersection of $P(\ba)$ with the $\ba$-axis will be close to $\ba=1$,
therefore in Eq.~(\ref{pbeta}) both terms are well approximated by a
Taylor expansion around $\ba=1$ up to linear order in $\ba$.

\section{Discussion}
We may conclude first of all that for the Lorentz gas the calculation of
corrections to the topological pressure due to open boundary conditions is
remarkably simple; it just requires the solution of an effective
$\ba$-dependent diffusion equation with the same open boundary conditions.
One question that could be asked here is whether the diffusion coefficient
to be used in this calculation is the same as in a closed system. Since
realizations of the random flight process with a slightly higher than
average collision frequency will have a slightly lower diffusion
coefficient these will lead to a slightly smaller escape rate from the
system.  The same can be argued for processes in which the frequency of
backscattering events is slightly enhanced and that of forward scattering
is slightly suppressed with respect to the averages in an equilibrum
system. Therefore the average diffusion coefficient on the repeller should
be slightly smaller than the diffusion coefficent of the equilibrium
system. However, one easily estimates that the suppression of the
diffusion coefficient is of order $k^2_0$, leading to a correction of
order $k_0^4$ in the escape rate. Our results for the corrections to
entropies and dimensions, which are all of order $k_0^2$, therefore will
not be changed. For higher order corrections of course such terms are
important. 

It is very interesting that for the dilute random Lorentz gas the effects
of the open boundary can be separated straightforwardly into effects due
to changes in the distributions of free flight times and of scattering
angles respectively. This is because the stretching factor first of all
factorizes (to leading order in the density) into a product of stretching
factors pertaining to a single free flight plus subsequent collision, and
in addition each of those factorizes into factors depending on the free
flight time respectively the scattering angles alone. Specifically, one
has $\Lambda_d(\theta,t)=\Lambda_1\Lambda_2$, with
$\Lambda_1=(2vt/a)^{d-1}$ and $ \Lambda_2=\cos^{d-3}\theta$. Rewriting the
factor $\Lambda^{1-\beta}$ as $\Lambda_1^{1-\beta_1}\Lambda_2^{1-\beta_2}$
one may obtain the topological pressure and the effective diffusion
constant as functions of both $\ba_1$ and $\ba_2$. By taking the
derivatives of the topological pressure with respect to $\ba_1$ and
$\ba_2$ one finds that the two terms on the right hand side of Eq.\
(\ref{hks0}) may be assigned to the distributions of free times and of
scattering angles respectively.  Similarly the correction to the sum of
the positive Lyapunov exponents, given by $k^2_0D'(1)$, can be separated
into a term due to the change in the free time distribution, which is of
the form
\be
\Delta_{t}\Sigma\lambda^+=k^2_0D(1)(d-1)\left[\ln \frac{2v}{\nu_d a}-{\cal C} -\frac{d-1}{d+1}\right],
\label{deltat}
\ee
and a term due to the change in the distribution of scattering angles,
reading
\be
\Delta_{
\theta}\Sigma\lambda^+=k^2_0\frac{(3-d) }2 D(1)\left[{\cal C} +
\Psi\left(\frac{d+1}2\right)-\frac {d-1}{d+1}\right].
\ee
One might be tempted to think that the changes in the distribution
function are due primarily to particles near the open boundary, which will
only survive if their free flights keep them inside the system. This
however is completely false. At any given instant the fraction of
particles inside a layer of a few mean free paths near the boundary may be
estimated as proportional to $1/R^2$, with $R$ an estimate for the
diameter of the system; the volume fraction covered by the boundary layer
is proportional to $1/R$ and the density near the open boundary is also of
order $1/R$ compared to the average density. In addition, in order for a
particle trajectory to be on the repeller, it has to remain inside the
system forever after. For trajectories near the open boundary at the given
instant, the probability for this to happen is another order $1/R$ smaller
than for trajectories at large~\cite{vbld2000}. Therefore trajectories
getting near the boundary at any time do not contribute at all to the
order $k^2_0$. Rather, the deviations from the equilibrium distributions
are caused by the fact that free flights or scattering angles that move a
particle away from an open boundary lead to higher survival probability
than ones that bring it closer to it. If one wants to know how, locally,
the distribution of free times or scattering angles is changed, he would
have to take recourse to the methods laid out in Ref.~\cite{vbld2000}. It
is fortunate though, that the reduction of the dynamical partition
function to an effective diffusion problem allows one to calculate the
KS-entropy and the like, at least up to order $k^2_0$, without having to
go through the details of this formalism.

An interesting question is in how far the present method of calculating
corrections to the topological pressure from the solution of a diffusion
equation can be generalized to systems of many moving particles.
Unfortunately this does not seem to be possible in any straightforward
way, even for dilute systems with hard interactions. The reason for this
is the semi-dispersing character of the dynamics of these systems,
interpreted as billiards in a high dimensional phase space. Due to this
property the dynamical partition function does not approximately factorize
into a product of terms describing the effects of one free flight and the
subsequent collision. Note that even the Lorentz gas loses this property
at higher densities, where the mean free path between collisions is not
large compared to the scatterer radius.

Finally, we remark that, like in the equilibrium and field driven
disordered Lorentz gas, the calculations of the topological pressure for
general values of $\ba$ have to be taken with caution
\cite{avbed,vbd,mvb2004}. Points in phase space are weighted by
$\Lambda^{1-\ba}$, thus for $\ba$ sufficiently $<1$ the dynamical
partition function will be dominated by the largest stretching factors,
which are due to trajectories confined to regions of high densities of
scatterers.  Even though the number of such trajectories decreases
exponentially with time, the stretching factors  raised to the power
$1-\ba$ of the remaining ones will still make these dominant for $\ba$
sufficiently far from 1. For $\beta$ sufficiently $>1$ on the other hand
the dynamical partition function will be dominated by trajectories
confined to the neighborhood of the least unstable periodic orbit.

In conclusion, we have shown how to relate the thermodynamic formalism for
the open Lorentz gas to a diffusive random flight problem. We have
calculated the topological pressure in $d$ dimensions as a function of the
temperature-like parameter $\ba$. For the open Lorentz gas, the
topological pressure is the sum of the topological pressure for the closed
system and a $\ba$-dependent effective escape rate which is given by a
$\ba$-dependent diffusion coefficient multiplied by the square of the
smallest wave number $k_0$ fitting the diffusion equation with absorbing
boundary equations. From this we have obtained several dynamical
quantities such as the Kolmogorov-Sinai entropy, the topological entropy,
and the partial information dimension for general dimension $d$. 

This work was supported by the {\sl Collective and cooperative statistical
physics phenomena} program of FOM (Fundamenteel Onderzoek der Materie).


\begin{thebibliography}{18}
\expandafter\ifx\csname natexlab\endcsname\relax\def\natexlab#1{#1}\fi
\expandafter\ifx\csname bibnamefont\endcsname\relax
  \def\bibnamefont#1{#1}\fi
\expandafter\ifx\csname bibfnamefont\endcsname\relax
  \def\bibfnamefont#1{#1}\fi
\expandafter\ifx\csname citenamefont\endcsname\relax
  \def\citenamefont#1{#1}\fi
\expandafter\ifx\csname url\endcsname\relax
  \def\url#1{\texttt{#1}}\fi
\expandafter\ifx\csname urlprefix\endcsname\relax\def\urlprefix{URL }\fi
\providecommand{\bibinfo}[2]{#2}
\providecommand{\eprint}[2][]{\url{#2}}

\bibitem[{\citenamefont{Sinai}(1972)}]{sinai}
\bibinfo{author}{\bibfnamefont{Y.~G.} \bibnamefont{Sinai}},
  \bibinfo{journal}{Russ.\ Math.\ Surv.} \textbf{\bibinfo{volume}{27}},
  \bibinfo{pages}{21} (\bibinfo{year}{1972}).

\bibitem[{\citenamefont{Ruelle}(1978)}]{ruelle}
\bibinfo{author}{\bibfnamefont{D.}~\bibnamefont{Ruelle}},
  \emph{\bibinfo{title}{Thermodynamic Formalism}}
  (\bibinfo{publisher}{Addison-Wesley Publishing Co., New York},
  \bibinfo{year}{1978}).

\bibitem[{\citenamefont{Bowen}(1975)}]{bowen}
\bibinfo{author}{\bibfnamefont{R.}~\bibnamefont{Bowen}}, in
  \emph{\bibinfo{booktitle}{Lecture Notes in Mathematics}}
  (\bibinfo{publisher}{Springer Verlag, Berlin}, \bibinfo{year}{1975}), vol.\
  \bibinfo{volume}{470}.

\bibitem[{\citenamefont{Eckmann and Ruelle}(1985)}]{er1985}
\bibinfo{author}{\bibfnamefont{J.-P.} \bibnamefont{Eckmann}} \bibnamefont{and}
  \bibinfo{author}{\bibfnamefont{D.}~\bibnamefont{Ruelle}},
  \bibinfo{journal}{Rev.\ Mod.\ Phys.} \textbf{\bibinfo{volume}{57}},
  \bibinfo{pages}{617} (\bibinfo{year}{1985}).

\bibitem[{\citenamefont{Beck and Schl{\"o}gl}(1993)}]{beck}
\bibinfo{author}{\bibfnamefont{C.}~\bibnamefont{Beck}} \bibnamefont{and}
  \bibinfo{author}{\bibfnamefont{F.}~\bibnamefont{Schl{\"o}gl}},
  \emph{\bibinfo{title}{Thermodynamics of chaotic systems}}
  (\bibinfo{publisher}{Cambridge University Press, New York},
  \bibinfo{year}{1993}).
  
\bibitem[{\citenamefont{Sinai}(1970)}]{SB}
\bibinfo{author}{\bibfnamefont{Y.~G.} \bibnamefont{Sinai}},
  \bibinfo{journal}{Russ.\ Math.\ Surv.} \textbf{\bibinfo{volume}{25}},
  \bibinfo{pages}{137} (\bibinfo{year}{1970}).



\bibitem[{\citenamefont{Hauge}(1999)}]{Hauge}
\bibinfo{author}{\bibfnamefont{E.~H.} \bibnamefont{Hauge}},
\bibinfo{title}{\emph{"What can one learn from Lorentz Models?"} in \emph {Transport Phenomena} edited by G.\ Kirczenow and J.\ Marro, Lecture notes in Physics} (\bibinfo{publisher}{Springer Verlag, Berlin},
  \bibinfo{year}{1999}), vol.\ \bibinfo{volume}{31}, \bibinfo{pages}{337}.


\bibitem[{\citenamefont{Dorfman}(1999)}]{dorfman}
\bibinfo{author}{\bibfnamefont{J.~R.} \bibnamefont{Dorfman}},
  \emph{\bibinfo{title}{An Introduction to Chaos in Non-Equilibrium Statistical
  Mechanics}} (\bibinfo{publisher}{Cambridge University Press, New York},
  \bibinfo{year}{1999}).

\bibitem[{\citenamefont{van Beijeren}(1982)}]{vb1982}
\bibinfo{author}{\bibfnamefont{H.}~\bibnamefont{van Beijeren}},
  \bibinfo{journal}{Rev.\ Mod.\ Phys.} \textbf{\bibinfo{volume}{54}},
  \bibinfo{pages}{195} (\bibinfo{year}{1982}).


\bibitem[{\citenamefont{van Leeuwen and Weijland}(1982)}]{vLW}
\bibinfo{author}{\bibfnamefont{J~ M.~J.\ }\bibnamefont{van Leeuwen}},
  \bibnamefont{and} \bibinfo{author}{\bibfnamefont{A.\ } \bibnamefont{Weyland}}, 
  \bibinfo{journal}{Phys.\ } \textbf{\bibinfo{volume}{36}},
  \bibinfo{pages}{457} (\bibinfo{year}{1967}).


\bibitem[{\citenamefont{Weyland and van Leeuwen}(1982)}]{WvL}
\bibinfo{author}{\bibfnamefont{A.\ }\bibnamefont{Weyland}},
  \bibnamefont{and}  \bibinfo{author}
 {\bibfnamefont{J.~M.~J.\ }\bibnamefont{van Leeuwen}},
  \bibinfo{journal}{Phys.\ } \textbf{\bibinfo{volume}{38}},
  \bibinfo{pages}{35} (\bibinfo{year}{1968}).

\bibitem[{\citenamefont{Ernst and Weyland}(1982)}]{EW}
\bibinfo{author}{\bibfnamefont{M.~H.\ }\bibnamefont{Ernst}},
  \bibnamefont{and} \bibinfo{author}{\bibfnamefont{A.\ }
  \bibnamefont{Weyland}}, 
  \bibinfo{journal}{Phys.~Lett.~A} \textbf{\bibinfo{volume}{34}},
  \bibinfo{pages}{39} (\bibinfo{year}{1971}).
  
\bibitem[{\citenamefont{Lorentz}(1905)}]{lorentz}
\bibinfo{author}{\bibfnamefont{H.~A.} \bibnamefont{Lorentz}},
  \bibinfo{journal}{Proc.~Roy.\ Acad.\ Amst.} \textbf{\bibinfo{volume}{7}},
  \bibinfo{pages}{438} (\bibinfo{year}{1905}).

\bibitem[{\citenamefont{van Beijeren and Dorfman}(2002)}]{vbd}
\bibinfo{author}{\bibfnamefont{H.}~\bibnamefont{van Beijeren}}
  \bibnamefont{and} \bibinfo{author}{\bibfnamefont{J.~R.}
  \bibnamefont{Dorfman}}, \bibinfo{journal}{J.\ Stat.\ Phys.\ }
  \textbf{\bibinfo{volume}{108}}, \bibinfo{pages}{767} (\bibinfo{year}{2002}).

\bibitem[{\citenamefont{van Beijeren et~al.}(1998)\citenamefont{van Beijeren,
  Latz, and Dorfman}}]{vbld}
\bibinfo{author}{\bibfnamefont{H.}~\bibnamefont{van Beijeren}},
  \bibinfo{author}{\bibfnamefont{A.}~\bibnamefont{Latz}}, \bibnamefont{and}
  \bibinfo{author}{\bibfnamefont{J.~R.} \bibnamefont{Dorfman}},
  \bibinfo{journal}{Phys.\ Rev.\ E} \textbf{\bibinfo{volume}{57}},
  \bibinfo{pages}{4077} (\bibinfo{year}{1998}).

\bibitem[{\citenamefont{van Beijeren et~al.}(2000)\citenamefont{van Beijeren,
  Latz, and Dorfman}}]{vbld2000}
\bibinfo{author}{\bibfnamefont{H.}~\bibnamefont{van Beijeren}},
  \bibinfo{author}{\bibfnamefont{A.}~\bibnamefont{Latz}}, \bibnamefont{and}
  \bibinfo{author}{\bibfnamefont{J.~R.} \bibnamefont{Dorfman}},
  \bibinfo{journal}{Phys.\ Rev.\ E} \textbf{\bibinfo{volume}{63}},
  \bibinfo{pages}{016312} (\bibinfo{year}{2000}).

\bibitem[{\citenamefont{de~Wijn and van Beijeren}(2004)}]{dwvb2004}
\bibinfo{author}{\bibfnamefont{A.~S.} \bibnamefont{de~Wijn}} \bibnamefont{and}
  \bibinfo{author}{\bibfnamefont{H.}~\bibnamefont{van Beijeren}},
  \bibinfo{journal}{Phys.\ Rev.\ E} \textbf{\bibinfo{volume}{70}},
  \bibinfo{pages}{036209} (\bibinfo{year}{2004}).

\bibitem[{\citenamefont{van Beijeren and Dorfman}(1995)}]{vbd1995}
\bibinfo{author}{\bibfnamefont{H.}~\bibnamefont{van Beijeren}}
  \bibnamefont{and} \bibinfo{author}{\bibfnamefont{J.~R.}
  \bibnamefont{Dorfman}}, \bibinfo{journal}{Phys.\ Rev.\ Lett.}
  \textbf{\bibinfo{volume}{74}}, \bibinfo{pages}{4412} (\bibinfo{year}{1995}).

\bibitem[{\citenamefont{M{\"u}lken and van Beijeren}(2004)}]{mvb2004}
\bibinfo{author}{\bibfnamefont{O.}~\bibnamefont{M{\"u}lken}} \bibnamefont{and}
  \bibinfo{author}{\bibfnamefont{H.}~\bibnamefont{van Beijeren}},
  \bibinfo{journal}{Phys.\ Rev.\ E} \textbf{\bibinfo{volume}{69}},
  \bibinfo{pages}{046203} (\bibinfo{year}{2004}).

\bibitem[{\citenamefont{Gradshteyn and Ryzik}(1965)}]{gradshteyn}
\bibinfo{author}{\bibfnamefont{I.~S.} \bibnamefont{Gradshteyn}}
  \bibnamefont{and} \bibinfo{author}{\bibfnamefont{I.~M.} \bibnamefont{Ryzik}},
  \emph{\bibinfo{title}{Tables of Integrals, Series, and Products}}
  (\bibinfo{publisher}{Academic Press, New York and London},
  \bibinfo{year}{1965}).

\bibitem[{\citenamefont{Gaspard and Nicolis}(1990)}]{gn1990}
\bibinfo{author}{\bibfnamefont{P.}~\bibnamefont{Gaspard}} \bibnamefont{and}
  \bibinfo{author}{\bibfnamefont{G.}~\bibnamefont{Nicolis}},
  \bibinfo{journal}{Phys.\ Rev.\ Lett.} \textbf{\bibinfo{volume}{65}},
  \bibinfo{pages}{1693} (\bibinfo{year}{1990}).

\bibitem[{\citenamefont{Bohr and Rand}(1987)}]{br1987}
\bibinfo{author}{\bibfnamefont{T.}~\bibnamefont{Bohr}} \bibnamefont{and}
  \bibinfo{author}{\bibfnamefont{D.}~\bibnamefont{Rand}},
  \bibinfo{journal}{Physica D} \textbf{\bibinfo{volume}{25}},
  \bibinfo{pages}{387} (\bibinfo{year}{1987}).

\bibitem[{\citenamefont{Appert et~al.}(1996)\citenamefont{Appert, van Beijeren,
  Ernst, and Dorfman}}]{avbed}
\bibinfo{author}{\bibfnamefont{C.}~\bibnamefont{Appert}},
  \bibinfo{author}{\bibfnamefont{H.}~\bibnamefont{van Beijeren}},
  \bibinfo{author}{\bibfnamefont{M.~H.} \bibnamefont{Ernst}}, \bibnamefont{and}
  \bibinfo{author}{\bibfnamefont{J.~R.} \bibnamefont{Dorfman}},
  \bibinfo{journal}{Phys.\ Rev.\ E} \textbf{\bibinfo{volume}{54}},
  \bibinfo{pages}{R1013} (\bibinfo{year}{1996}).

\end{thebibliography}
\end{document}